\documentclass[12pt]{article}
\usepackage{amssymb}
\usepackage{multirow}

\newcommand{\jpsi}{J\kern-0.1em/\kern-0.1em\psi\kern0.03em}

\begin{document}

\centerline{\Large \bf \boldmath
 Strange pentaquarks and excited \,$\Xi$\, hyperons}
\centerline{\Large \bf \boldmath
\vrule width 0pt height 3.5ex
in \,$\Xi_b^- \to \jpsi \Lambda K^-$\, final states%
\footnote{To appear in {\em 
Science Bulletin (2021), 
{\tt doi: https://doi.org/10.1016/j.scib.2021.04.013}
}}
}
\bigskip

\centerline{\vrule width 0pt height 4.5ex
Marek Karliner$^a$\footnote{{\tt marek@tauex.tau.ac.il}}
 and Jonathan L. Rosner$^b$\footnote{{\tt rosner@hep.uchicago.edu}}}
\medskip
\medskip

\centerline{$^a$ {\it School of Physics and Astronomy}}
\centerline{\it Tel Aviv University, Tel Aviv 69978, Israel}
\medskip

\centerline{$^b$ {\it Enrico Fermi Institute and Department of Physics}}
\centerline{\it University of Chicago, 5640 S. Ellis Avenue, Chicago, IL
60637, USA}
\bigskip
\strut

Besides the observed quark-antiquark and three-quark states, the quark model
predicts other states with integral charges, such as $qq \bar q \bar q$
(tetraquarks) and $qqqq \bar q$ (pentaquarks).  In 2015 the LHCb Collaboration
observed $\jpsi p$ resonances consistent with pentaquarks in $\Lambda_b^0 \to
\jpsi p K^-$ decays \cite{Aaij:2015tga},
with minimal quark content $c\bar c uud$.  Two states were seen:  a narrow one
at 4450 MeV and a broader one at 4380 MeV.  A subsequent analysis with more
data \cite{Aaij:2019vzc} resolved the peak at 4450 MeV into two, separated by
17 MeV, discovered another narrow one at 4312 MeV. 
Ref.~\cite{Aaij:2019vzc} employed a one-dimensional analysis of the 
$\jpsi p$ spectrum, which is not sensitive to a broad state like
$P_c(4380)$, so the final verdict on the latter will have to wait till an
amplitude analysis of the additional data.
The low-mass peak may be a $J=1/2~\Sigma_c \bar D$
$S$-wave molecule, while the two upper ones may be $J=1/2$ and 3/2 $\,S$-wave
molecules of $\Sigma_c \bar{D}^*$ separated by hyperfine splitting.  The
mass $M$ and width $\Gamma$ of these peaks are:
$$
M=4311.9\pm0.7^{+6.8}_{-0.6} {\rm~MeV},~~
  \Gamma =9.8\pm2.7^{+3.7}_{-4.5}{\rm~MeV}~,
$$
$$
M=4440.3\pm1.3^{+4.1}_{-4.7} {\rm~MeV},~~
  \Gamma=20.6\pm4.9^{+8.7}_{-10.1}{\rm~MeV}~,
$$
$$
M=4457.3\pm0.6^{+4.1}_{-1.7}{\rm~MeV},~~
  \Gamma =6.4\pm2.0^{+5.7}_{-1.9}{\rm~MeV}~.
$$
The $\Sigma_c^+~\bar{D}^{*0}$ threshold is $(2452.9\pm0.4) + (2006.85\pm0.05)
= 4459.8 \pm 0.4$ MeV,
while the $\Sigma_c^{++} \bar{D}^{*-}$ threshold is $4464.23 \pm 0.15$ MeV.

The LHCb Collaboration has now reported evidence for a 
structure in the $\jpsi \Lambda$
final state produced in the decay $\Xi^-_b \to \jpsi \Lambda K^-$
\cite{Aaij:2020gdg}, with minimal quark content $c \bar c u d s$.
A peak is found with mass $4458.8\pm2.9^{+4.7}
_{-1.1}$ MeV and width $17.3\pm6.5^{+8.0}_{-5.7}$ MeV. 
The signal significance is 3.1 $\sigma$.  The structure is also
consistent with two resonances, with masses $4454.9\pm2.7$ and $4467.8\pm3.7$
MeV and widths $7.5\pm9.7$ and $5.2\pm5.3$ MeV, respectively.

The two-peak hypothesis is consistent with a molecular interpretation based on
constituents $\Xi_c^0~({M} = 2470.90^{+0.22}_{-0.29}$ MeV) and
$\bar D^{*0}~({M} = 2006.85\pm0.05$ MeV), corresponding to a threshold
of 4477.75$^{+0.23}_{-0.29}$ MeV.  An $S$-wave molecule of the spin-1/2 charmed
baryon and the spin-1 anticharmed meson can have total spin 1/2 or 3/2.
In this case the hyperfine splitting is $\Delta E(HF,s) = 12.9\pm4.6$ MeV.

The two-peak hyperfine splitting between the nonstrange pentaquarks at 4440
and 4457 MeV is $\Delta E(HF,ns) = 17$ MeV, with an error which depends on
correlations but which is probably small compared to that on $\Delta E(HF,s)$.
The ratio of the two hyperfine splittings is
$\Delta E(HF,s)/\Delta E(HF,ns) \approx 0.76\pm 0.27$.
In a molecular picture this ratio
depends on the reduced mass of the two molecules and on the couplings of
the mesons generating the hyperfine interactions. To get a rough idea of
the expected ratio, we note that
the reduced masses are very close, 1104 and 1107 MeV, respectively
and that pions are the lightest mesons that can be exchanged in both
$\Sigma_c \bar{D}^*$  and $\Xi_c \bar{D}^*$.  
The ratio of the relevant couplings
\ $ R_{\Xi_c/\Sigma_c} \equiv g_{\Xi \Xi \pi}/g_{\Sigma_c \Sigma_c \pi} $ \ 
is not known, but since $\Xi_c$ has isospin $1/2$, 
while $\Sigma_c$ has $I{=}1$,
one expects $R_{\Xi_c/\Sigma_c}\lesssim {\cal O}(1)$,
so $\Delta E(HF,s)/\Delta E(HF,ns)$
is expected to be $\lesssim 1$, as observed.

The upshot is that new LHCb results suggest that in addition to the original
$\Sigma_c \bar D^*$ pentaquark hadronic molecule with quark content $c \bar c u
u d$ we now have very suggestive evidence for a $\Xi_c \bar D^*$ hadronic
molecule with quark content $c \bar c u d s$.

A somewhat subtle point is that the new $\Xi_c \bar D^*$ state 
{\em does not}  correspond to an $SU(3)_F$ rotation of the original
$\Sigma_c \bar D^*$ state. This is because under $SU(3)_F$ rotation
$\Sigma_c$ goes to $\Xi_c^\prime$, rather than to $\Xi_c$.
This implies that in addition to \,$\Xi_c \bar D^*$\,
a \,$\Xi_c^\prime \bar D^*$\, hadronic molecule might exist, 
with a mass shifted upwards by
approximately the $\Xi_c^\prime - \Xi_c$ mass difference, i.e., about 108 MeV.

Moreover, if the $\Xi_c \bar D^{*0}$ molecule consists of two peaks, 
as suggested by Fig.~6 of  Ref.~\cite{Aaij:2020gdg}, one with $J=1/2$ and the
other with $J=3/2$, the same should be true for the $\Xi_c^\prime \bar D^{*0}$
molecule.  With a little imagination, Fig. 3(right) of Ref.~\cite{Aaij:2020gdg}
could be consistent with that. Of course at this stage it is only a
conjecture.  Clearly more data are needed, but the possibility is intriguing.

The LHCb collaboration has also studied the excited $\Xi^-$ states decaying to
$\Lambda K^-$. (Notation is that of \cite{Zyla:2020zbs}.)  Strong signals are
seen for $\Xi^-(1690)$ and $\Xi^-(1820)$.  The latter is known to have
$J^P=3/2^-$,
but the former's $J^P$ is still unknown, with major contenders being $1/2^\pm$.
States seen at 1911 and 2023 MeV are compatible with $\Xi(1950)$ and
$\Xi(2030)$ of Ref.\ \cite{Zyla:2020zbs}.  It would be interesting for LHCb to
examine the excited $\Xi^{*0}\to\Xi^-\pi^+$ spectrum, as a state $\Xi^0(1620)$
reported by Belle \cite{Belle} has a mass in some tension with quark model
predictions, which are closer to 1690 MeV for $J=1/2$ and either parity.

The claimed $\Xi^0(1620)$ is very close to $\Lambda \bar{K}^0$ threshold (1613
MeV), so it is not surprising that it doesn't show up in $\Xi_b^- \to \jpsi 
\Lambda K^-$, but it should be looked for in $\Xi_b^0 \to \jpsi \Xi^- \pi^+$.
The appearance of the $\Xi^0(1620)$ in the $\Xi^- \pi^+$ channel could be an
effect of the opening of the $\Lambda \bar{K}^0$ 
channel, without being a genuine quark model state.

\section*{Acknowledgements}
The original publication is available at 
{\tt https://www.scichina.com}
\ and 
\hfill\break
{\tt https://www.journals.elsevier.com/science-bulletin},
\hfill\break
{\tt doi: https://doi.org/10.1016/j.scib.2021.04.013}\,.
\newline
The research of M.K. was supported in part by NSFC-ISF
grant No.\ 3423/19.

\end{document}